\documentclass[%
mathleft,%
]{an}
\usepackage{graphicx}
\usepackage{times}
\pdfoutput=1

\begin{document}


\title{Light curve solutions of the ultrashort-period \emph{Kepler} binaries\,\thanks{Data from \emph{Kepler}}}

\author{D. Kjurkchieva\inst{1}\fnmsep\thanks{Corresponding author:
  \email{d.kyurkchieva@shu-bg.net}\newline}
\and D. Dimitrov\inst{2} }
\titlerunning{Light curve solutions of the ultrashort-period \emph{Kepler} binaries}
\authorrunning{D. Kjurkchieva \& D. Dimitrov}
\institute{Department of Physics, Shumen University, 115 Universitetska, 9700 Shumen, Bulgaria \and 
Institute of Astronomy and NAO, Bulgarian Academy of Sciences, Tsarigradsko shossee 72, 1784 Sofia, Bulgaria}

\received{... 2014} \accepted{... 2014} \publonline{later}


\abstract{We carried out light curve solutions of the ultrashort-period binaries with MS components observed by \textit{Kepler}. All six targets 
turned out almost in thermal contact with contact or slightly overcontact configurations. Two of them, KID~4921906 and KID~6309193, are not eclipsing 
but reveal ellipsoidal and spot variability. One of the components of KID~8108785 exhibits inherent, quasi-sinusoidal, small-amplitude variability. 
KID~12055255 turned out a very rare case of ultrashort-period overcontact binary consisting of two M dwarfs. Our modeling indicated that the 
variability of KID~9532219 is due to eclipses but not to $\delta$ Sct pulsations as it was previously supposed.}

\maketitle

\section{Introduction}
Most of the contact binaries consisting of solar-type components have orbital periods within 0.25 d $< P < $ 0.7 d. Rucinski (1992) found that they 
show a short-period limit at about 0.22 d. But the modern large surveys allowed to discover targets beyond this limit (Maceroni \& Rucinski 1997; 
Maceroni \& Montalban 2004; Weldrake 2004; Rucinski 2006, 2007; Pribulla et al. 2009; Lohr et al. 2014; Qian et al. 2014, etc.).

The well-studied binaries with periods around and below cut-off limit (ultrashort-period systems) and MS components are quite rare but important 
objects for the astrophysics, especially for the stellar evolution.

Until two decades ago the statistics of contact binaries was particularly affected by the trend to discover relatively large light-curve amplitudes, 
but very recently, the huge surveys (ROTSE, MACHO, ASAS, SuperWASP, etc.) led to discoveries of many low-amplitude systems.

Lately, \textit{Kepler} provides unique, unprecedented precise, extended and nearly uninterrupted data set for a large number and variety of stars. 
Above two thousands eclipsing binaries (EBs) were identified and included in the \textit{Kepler} EB catalog (Prsa et al. 2011, further PRSA11; 
Slawson et al. 2011). According to the initial classification the sample consists of 1261 detached, 152 semidetached, 469 overcontact binaries, 
137 ellipsoidal variables (they are not EBs!), and 146 uncertain systems. The automated fitting of the light curves by a polynomial chain and the 
artificial intelligence based tier \textit{EBAI} (Prsa et al. 2008) were used for the morphological classification and determination of the global 
parameters of some systems of the \textit{Kepler} EB catalog.

Several individual \textit{Kepler} EBs were studied in details (e.g., Southworth et al. 2011; Steffen et al. 2011; Welsh et al. 2011; Winn et al. 
2011). This paper is devoted to standard ("manual") light curve solutions of the ultrashort-period binaries from the \textit{Kepler} EB catalog.

\section{Selection of targets and preliminary preparation of the data}

As a continuation of our interest in binaries with extremely small orbital periods (Dimitrov \& Kjurkchieva 2010), we reviewed the \textit{Kepler} EB 
catalog. It contains 22 targets with periods $P < 0.23$ d (for better statistics we put an upper limit 0.23 d that is slightly bigger than the 
cut-off limit of 0.22 d).

Although Prsa et al. (2011) reported about visual inspection of every target and culling about 27$\%$ of their initial sample as nonEBs 
(RR Lyr-, $\gamma$ Dor-, $\delta$-Sct type stars), we carried out new visual review of the chosen 22 targets. We concluded that 13 ultrashort-period 
stars are $\delta$ Sct variables due to the shape of their curves (KID 10417986, KID 8912468, KID 8758716, KID 10855535, KID 9612468, KID 8912730, 
KID 9898401, KID 7375612, KID 5872696, KID 10684673, KID 6699679, KID 6287172, KID 11825204). Our classification of these stars is supported by their 
high temperatures (above 6000 K).

One ultrashort-period star, KID~9472174, turned out to be the known sdB+dM binary 2M1938+4603.

The shapes of the light curves of two ultrashort-period targets, KID~8288741 and KID~12350008, seem as those of detached eclipsing systems 
(with almost flat maxima) but their amplitudes of 1--2 mmag and periods are absolutely inappropriate for such configurations. Our preliminary attempts 
to reproduce their data required enormous big third light. This means that KID~8288741 and KID~12350008 are not the true variable stars. It is 
meaningless to model these data because the obtained parameters would not be reliable. Matuevic et al. (2012) noted that the automated classification 
and modeling of \textit{Kepler} EBs, even with locally linear embedding, failed for small amplitudes ($\sim$0.001 mag). KID~8288741 and KID~12350008 
are such cases but the reason to be removed from the further analysis is the discrepancy between the light curve shape appropriate for detached 
eclipsing binary and the negligible amplitude of variability (smaller even than those of ellipsoidal variations). 

As a result of the selection procedure we gathered sample of six ultrashort-period binaries: KID 4921906, KID 1572353, KID 8108785, KID 6309193, 
KID 12055255, KID 9532219. More conclusive confirmation of their duplicity would be spectroscopic data. The objects would require 8m-class telescope 
due to low brightness and short period at the same time. Another possibility for proper classification is the color-period diagram (Duerbeck, 1997) 
but the presence of third component or close neighbor can alter the color information. So we will use the important criterion for sure classification 
of the targets: the successful reproducing of their light curves by effects of duplicity. 

Table 1 presents information for our 6 targets from the \textit{Kepler }EB catalog: \textit{Kepler} magnitude $m_{K}$; epoch $T_{0}$ of the primary 
minimum; orbital period $P$; mean temperature $T_{m}$; metallicity $[Fe/H]$; radius $R$; $\log g$; reddening E(B-V); type of configuration; ratio
$T_2/T_1$ of the temperatures of the components; fillout factor \textit{FF}; crowding (a number between 0 and 1 that specifies the fraction of flux 
in the optimal photometric aperture due to the target star with respect to the total flux from all sources).

\begin{table*}
\begin{minipage}[t]{\columnwidth}
\caption{Parameters of the targets from the EB catalog}
\label{tab:1} \centering
\begin{scriptsize}
\renewcommand{\footnoterule}{}
\begin{tabular}{ccccccccccccc}
\hline
Kepler ID&  $m_K$ & $T_0$-2400000 &   $P$  &  $T_m$ & [Fe/H] & $R$ & $\log{g}$ & E(B-V)& Type & $T_2/T_1$ & $FF$ & crowd \\
         &  [mag] &  [days]       & [days] &  [K]   &        & [$R_{\sun}$]&       &     &      &      &      &        \\
\hline

4921906 &  15.203  & 54965.148704 & 0.213732 & 4800 & -0.394 & 0.99 & 4.41 & 0.109 & OC & 0.958 & 1.033 & 0.648  \\

1572353 &  15.234  & 55001.306290 & 0.228900 & 4470 & -0.261 & 0.77 & 4.54 & 0.076 & OC & 0.883 & 1.051 & 0.9    \\

8108785 &  14.757  & 54964.641700 & 0.228840 & 4354 & -0.298 & 0.67 & 4.61 & 0.053 & OC & 0.855 & 0.414 & 0.921   \\

6309193 &  13.674  & 55002.434080 & 0.212500 & 5367 & -0.383 & 1.09 & 4.38 & 0.098 & OC & 1.012 & 1.01  & 0.92   \\

12055255&  15.866  & 54964.966610 & 0.220950 & 4093 &  0.106 & 0.73 & 4.50 & 0.066 & ?  &   ?   &  ?    & 0.566  \\

9532219 &  16.118  & 55001.945246 & 0.198155 & 5031 & -0.539 & 1.06 & 4.38 & 0.159 & ?  &   ?   &   ?   & ?       \\
 \hline
\end{tabular}
\end{scriptsize}
\end{minipage}
\end{table*}

More than 50000 points are available for each target in the \textit{Kepler} archive. The detailed review of these data revealed that the shape of 
the light curves almost did not change during the different observational quarters. However the levels of maxima and minima and the amplitudes of 
the light curves of the most targets changed considerably during the different quarters causing the big thickness of the total light curves 
(from all quarters). This effect is probably due to problems of the satellite steering and automated reduction of the data. That is why we assumed 
that it is reasonable to model data from several consecutive quarters rather than all data simultaneously. Moreover, we established that it is more 
appropriate to use the raw \textit{Kepler} data and to make own reduction and de-trending (Dimitrov et al. 2012).

We chose to model the data from quarters Q0 and Q1 for all targets excepting KID~9532219 which observations began later. Their folded light curves 
are presented in Figs. 1-6.

\begin{figure}
 \centering
 \includegraphics[width=0.88\columnwidth,keepaspectratio=true]{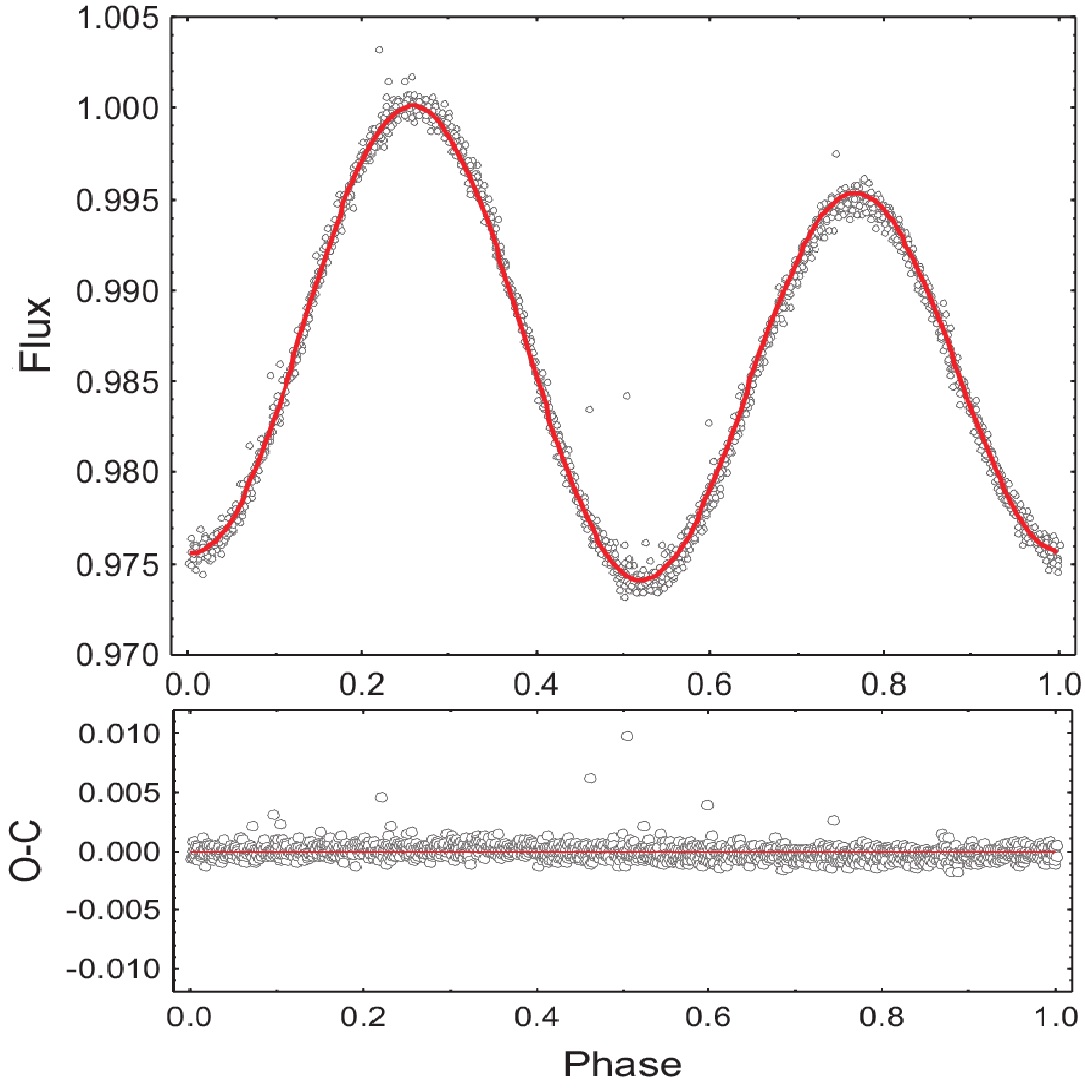}
 \caption{The light curve of KID~4921906 and our fit (top) and the corresponding residuals (bottom)}
 \label{fig:1}
\end{figure}

\begin{figure}
 \centering
 \includegraphics[width=0.85\columnwidth,keepaspectratio=true]{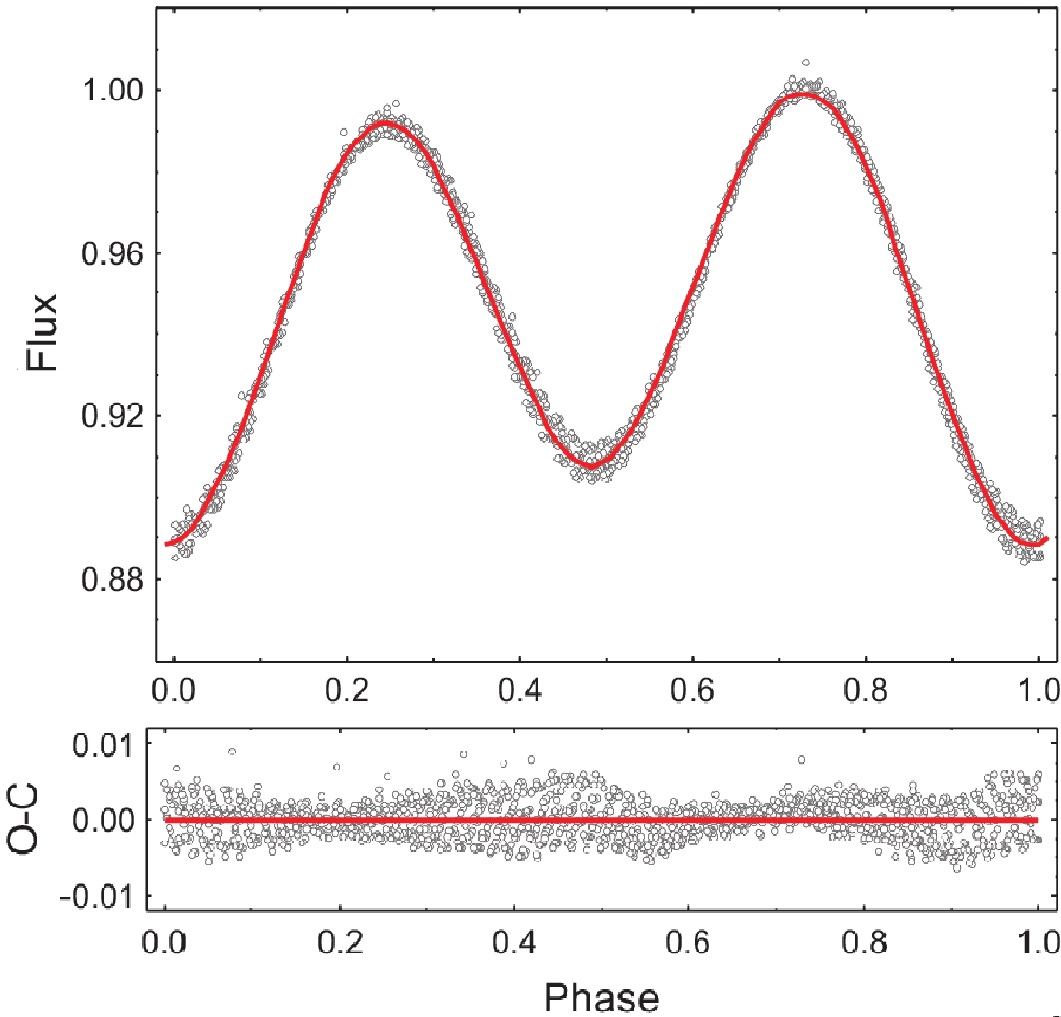}
 \caption{The light curve of KID~1572353 and our fit (top) and the corresponding residuals (bottom)}
 \label{fig:2}
\end{figure}

\begin{figure}
 \centering
 \includegraphics[width=0.85\columnwidth,keepaspectratio=true]{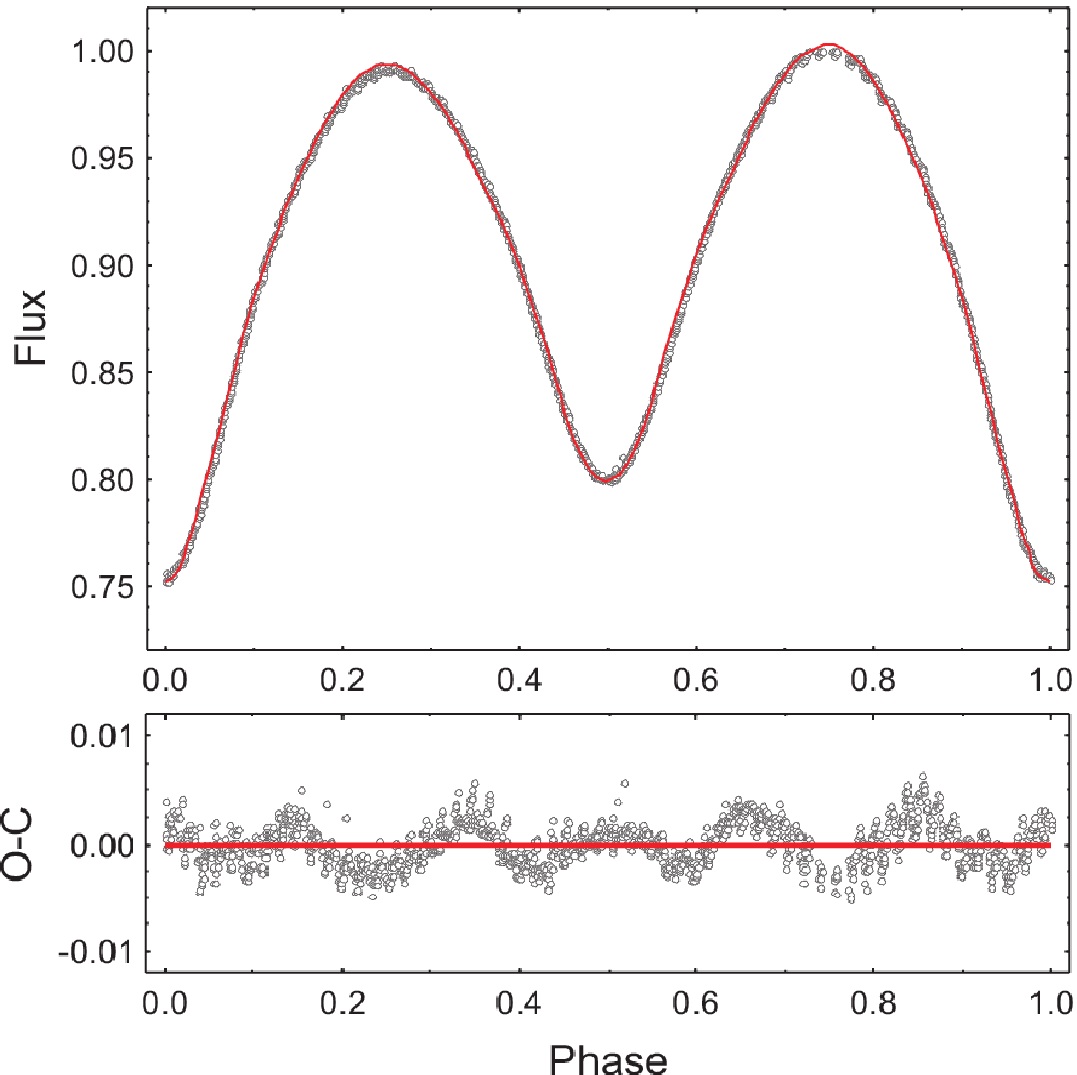}
 \caption{The light curve of KID~8108785 and our fit (top) and the corresponding residuals (bottom)}
 \label{fig:3}
\end{figure}

\begin{figure}
 \centering
 \includegraphics[width=0.84\columnwidth,keepaspectratio=true]{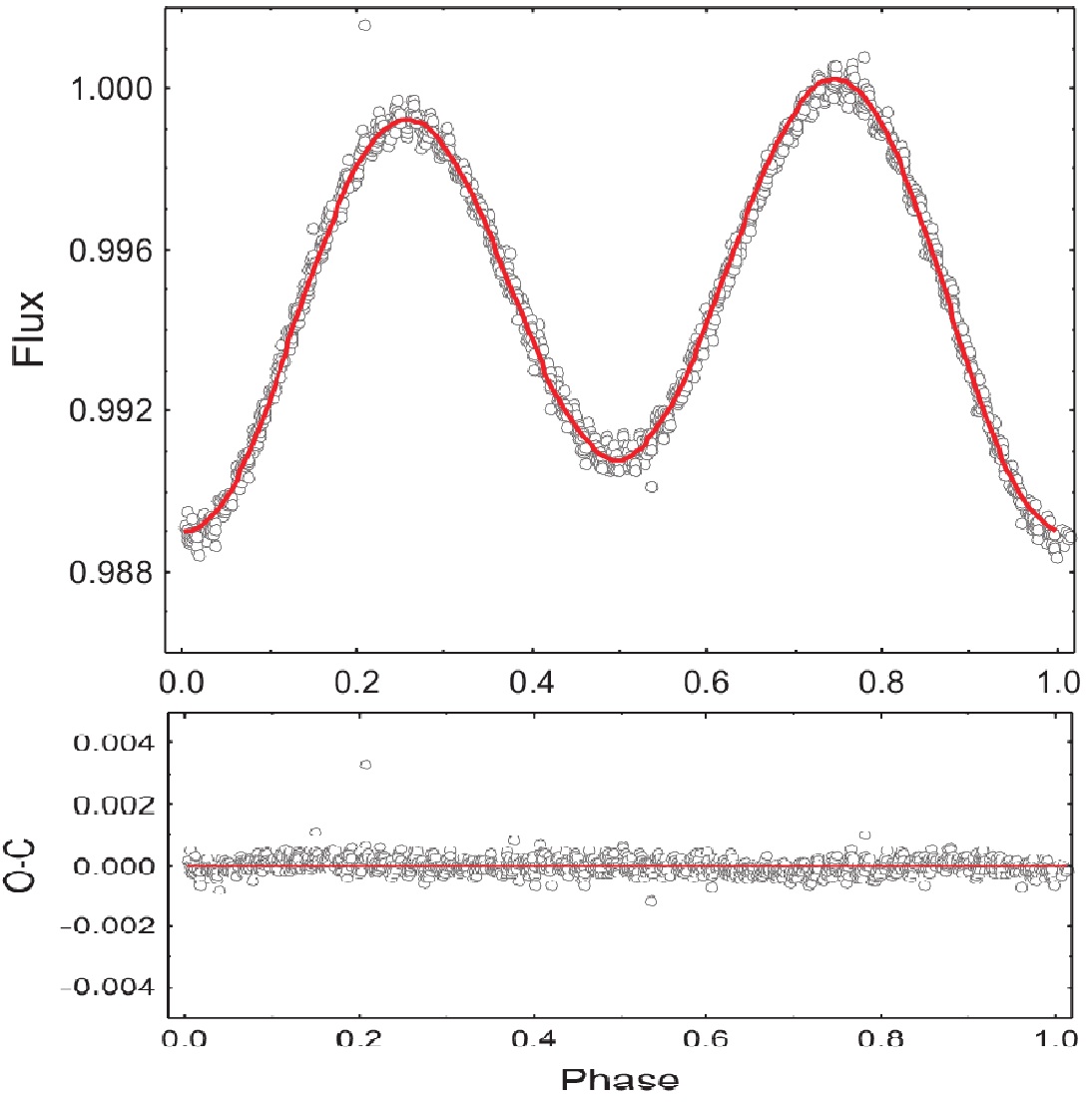}
 \caption{The light curve of KID~6309193 and our fit (top) and the corresponding residuals (bottom)}
 \label{fig:4}
\end{figure}

\begin{figure}
 \centering
 \includegraphics[width=0.85\columnwidth,keepaspectratio=true]{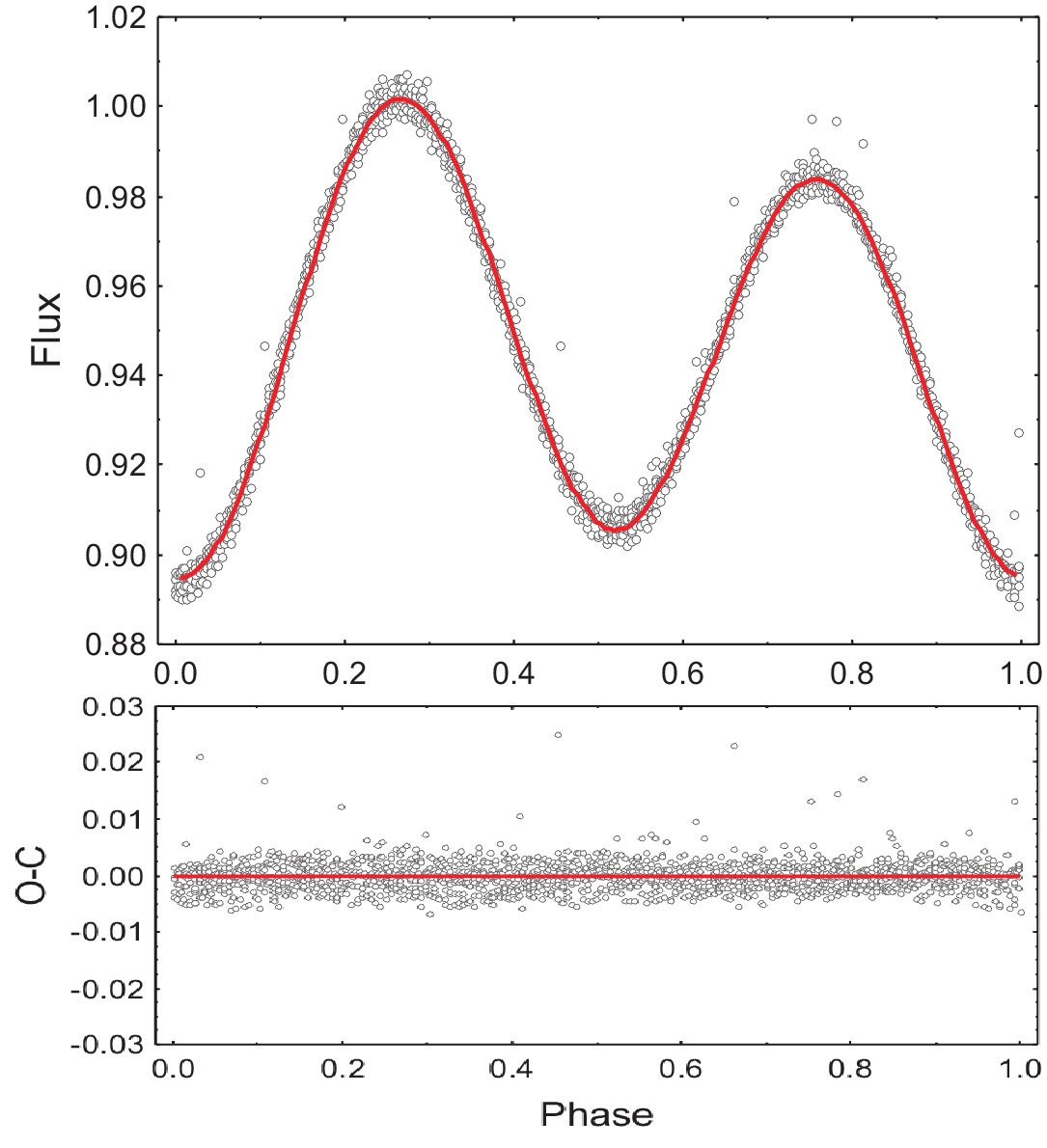}
 \caption{The light curve of KID~12055255 and our fit (top) and the corresponding residuals (bottom)}
 \label{fig:5}
\end{figure}

\begin{figure}
 \centering
 \includegraphics[width=0.9\columnwidth,keepaspectratio=true]{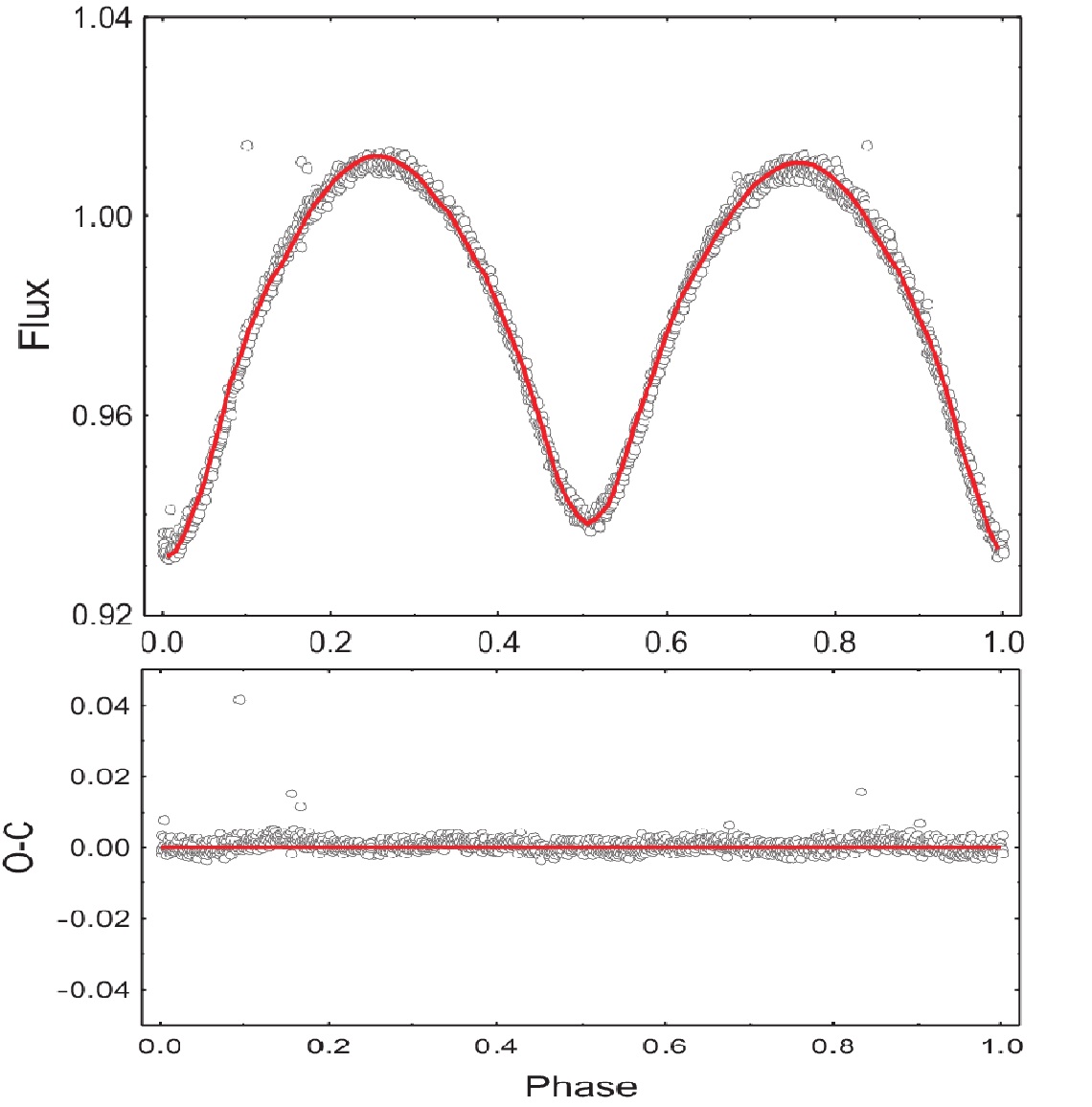}
 \caption{The light curve of KID~9532219 and our fit (top) and the corresponding residuals (bottom)}
 \label{fig:6}
\end{figure}

\section{Light curve solutions}

The shapes of the light curves (Figs. 1-6) implied that our targets are nearly contact or overcontact systems that was expected for their short 
orbital periods. The depths of the two minima of the most curves are close that means that the binaries are almost in thermal contact. The small 
light amplitudes of some targets point at low orbital inclinations and consequently, considerable contributions of the ellipsoidal variations.

Taking into account the foregoing considerations we carried out modeling of the photometric data by the code \textit{PHOEBE} (Prsa \& Zwitter 2005) 
using the following procedure.

Initially the primary temperature was fixed as $T_{1}$=$T_{m}$. The mean target temperatures $T_{m}$ (Table 1) have been estimated using dedicated 
pre-launch ground-based optical multi-color photometry plus Two Micron All Sky Survey (2MASS) $J$, $K$, and $H$ magnitudes (Skrutskie et al., 2006).

We varied the secondary temperature $T_{2}$, mass ratio $q$, orbital inclination $i$ and potentials $\Omega_{1,2}$ (and simultaneously relative 
radii $r_{1,2}$ and fillout factors $FF_{1,2}$). In order to reproduce the O'Connell effect we added cool spots on the primary and varied their 
parameters: longitude $\beta$, latitude $\lambda$, angular size $\alpha$ and temperature factor $k=T_{sp}/T_{1}$.

We adopted coefficients of gravity brightening 0.32 and reflection effect 0.5 appropriate for late stars. The limb-darkening coefficients were 
chosen according to the tables of Van Hamme (1993).

Finally, we varied the component temperatures around $T_{m}$ to search for the best fit.

The derived parameters of the targets corresponding to our light curve solutions are given in Table 2 while Table 3 presents information about the 
spot parameters. The formal \textit{PHOEBE} errors of the fitted parameters are quite small (Tables 2-3) due to the high precision of the 
\textit{Kepler} data.

It should be noted that the ephemerides from the \textit{Kepler} EB catalogue (PRSA11) did not lead to the best fits. For this aim we
varied the parameter ''shift'' of \textit{PHOEBE} which final values are given in column 2 of Table 2.

The synthetic curves corresponding to the parameters of our light curve solutions are shown in Figs. 1-6 as red continuous lines.

\begin{table*}
\begin{minipage}[t]{\columnwidth}
\caption{Parameters of the best light curve solutions}
\label{tab:2} \centering
\begin{scriptsize}
\renewcommand{\footnoterule}{}
\begin{tabular}{cccccccccccc}
\hline
KID     & shift   & $i$      & $q$ & $T_{1,2}$ & $T_{2}/T_{1}$ & $\Omega_{1,2}$ & FF$_{1,2}$ & $r^{mean}_{1,2}$ & $l_{1}$ & config & variability \\
        &         & [$\degr$]&     & [K]       &               &                &            &                  &         &        &             \\
\hline
4921906 & 0.0126  & 21.04$\pm$0.02 & 0.5977$\pm$0.0005 & 4528$\pm$14 & 0.960  & 3.0634$\pm$0.0002 & 0.0   & 0.424 & 0.66 & C      & ellipsoidal  \\
        &         &                &                   & 4348$\pm$12 &        & 3.0634$\pm$0.0002 & 0.0   & 0.334 &      &        &              \\
        &         &(26,4)          &(0,47)             &             & (0.958)&                   &       &       &      & (OC)   &              \\
1572353 & -0.01923& 43.84$\pm$0.02 & 0.9577$\pm$0.0003 & 4525$\pm$20 & 0.945  & 3.688$\pm$0.0003  & 0.0   & 0.383 & 0.58 & C      & grazing      \\
        &         &                &                   & 4277$\pm$13 &        & 3.688$\pm$0.001   & 0.0   & 0.374 &      &        & eclipses     \\
        &         & (33)           &(0.29)             &             & (0.883)&                   &       &       &      & (OC)   &              \\
8108785 & 0.00147 & 61.78$\pm$0.02 & 0.8309$\pm$0.0003 & 4711$\pm$4  & 0.901  & 3.4664$\pm$0.0004 & 0.0   & 0.395 & 0.62 & C      & eclipses     \\
        &         &                &                   & 4246$\pm$3  &        & 3.4664$\pm$0.0006 & 0.0   & 0.360 &      &        &              \\
        &         & (69.9)         &(0.33)             &             & (0.855)&                   &       &       &      & (OC)   &              \\
6309193 & -0.0196 &13.445$\pm$0.014& 0.8007$\pm$0.0003 & 5435$\pm$2.5& 0.874  & 3.4165$\pm$0.0005 & 0.0   & 0.398 & 0.68 & C      & ellipsoidal  \\
        &         &                &                   & 4751$\pm$2  &        & 3.4165$\pm$0.0005 & 0.0   & 0.359 &      &        &              \\
        &         & (21)           &(0.60)             &             & (1.012)&                   &       &       &      & (OC)   &              \\
12055255& 0.0128  & 39.79$\pm$0.05 & 0.9637$\pm$0.0005 & 3701$\pm$8  & 1.011  & 3.6326$\pm$0.0022 & 0.128 & 0.392 & 0.55 & OC     & grazing      \\
        &         &                &                   & 3744$\pm$10 &        & 3.6936$\pm$0.0012 & 0.019 & 0.377 &      &        & eclipses     \\
        &         & (?)            & (?)               &             & (?)    &                   &       &       &      & (?)    &              \\
9532219 & 0.00647 & 68.42$\pm$0.06 & 0.9320$\pm$0.0022 & 4950$\pm$22 & 0.983  & 3.5687$\pm$0.0022 & 0.172 & 0.398 & 0.55 & OC     & eclipses     \\
        &         &                &                   & 4867$\pm$21 &        & 3.5873$\pm$0.0028 & 0.164 & 0.388 &      &        &              \\
        &         & (?)            & (?)               &             & (?)    &                   &       &       &      & (?)    &              \\
\hline
\end{tabular}
\end{scriptsize}
Note: C means contact binaries, OC - overcontact systems
\end{minipage}

\end{table*}

\begin{table}
\begin{minipage}[t]{\columnwidth}
\caption{Parameters of the spots and third light of the targets}
\label{tab:3} \centering
\begin{scriptsize}
\renewcommand{\footnoterule}{}
\begin{tabular}{rccccc}
\hline
KID     & $\beta$ & $\lambda$ & $\alpha$ & $k$ & $l_3$    \\
        & [\degr]   & [\degr]     & [\degr]    &     &          \\
\hline
4921906 & 75.0$\pm$0.1  & 88.1$\pm$0.1  & 11.0$\pm$0.1 & 0.901$\pm$0.001 & 0 \\
1572353 & 91.5$\pm$0.1  & 329.9$\pm$0.1 & 17.6$\pm$0.1 & 0.900$\pm$0.002 & 0 \\
8108785 & 126.3$\pm$0.2 & 279.6$\pm$0.3 & 24.9$\pm$0.2 & 0.939$\pm$0.001 & 0 \\
6309193 & 63.9$\pm$0.1  & 338.4$\pm$0.1 & 10.3$\pm$0.2 & 0.940$\pm$0.001 & 0 \\
12055255& 59.6$\pm$0.2  & 89.9$\pm$0.2  & 14.9$\pm$0.2 & 0.900$\pm$0.001 & 0 \\
9532219 & 81.5$\pm$0.6  & 70.09$\pm$0.04& 10.1$\pm$0.7 & 0.914$\pm$0.007 & 0.78 \\
\hline
\end{tabular}
\end{scriptsize}
\end{minipage}
\end{table}

The derived photometric mass ratio requires a special attention. Usually this parameter needs spectral observations. But the spectral mass ratio for 
short-period binaries with fast rotation is quite imprecise due to the line blending while the geometrical effects and mass ratio are much more 
important in defining the shapes of their light curves and the eclipse depths. Hence, the values of the photometric mass ratio for ultrashort-period
\textit{Kepler} binaries should be considered with high confidence.

There are differences between our solutions and those obtained by automated fitting cited in the \textit{Kepler} EB catalog (the values in 
parentheses in Table 2).

\begin{description}
\item [(1)] The most orbital inclinations in PRSA11 are bigger than ours.
\item [(2)] The mass ratios in PRSA11 are considerably smaller than our values (Table 2).
\item [(3)] We had to assume a third light $l_{3}$ only for KID~9532219 (Table 3) while the values of this parameter in PRSA11 were nonzeros for the 
all six targets. We found a weaker star close to KID~9532219 (at distance around 6 arcsec) that could explain the presence of $l_{3}$ in the light 
curve solution of this target.
\item [(4)] There are no values of the potentials $\Omega_{1,2}$ (and thus of the relative radii $r_{1,2}$ and fillout factors $FF_{1,2}$) as
well as of relative luminosities in PRSA11 that means that the automated modelling leads to approximate results.
\end{description} 

There are several explanations for the differences between our "manual" solutions and those of the automated fitting.

\begin{description}
\item[(a)] The \textit{EBAI} method fixes $l_{3}$ = 1 - $crowding$ while this parameter is invoked in our procedure only when a satisfactory fit is 
impossible for any combination of the parameters. The bigger value of $l_{3}$ requires bigger light amplitude and correspondingly bigger orbital 
inclination. Moreover, due to the correlation between third light and mass ratio for partially eclipsing binaries the bigger third light
leads to the larger mass ratio for fixed inclination angle.
\item[(b)] Although all light curves exhibit O'Connell effect (Figs. 1-6) the \textit{EBAI} method does not take into account the effect from
spots. Then the results of the automated modelling should be assumed as first approximations of the fitted parameters.
\end{description}

\section{Analysis of the results}

The analysis of our light curve solutions of the ultrashort-period binaries from the \textit{Kepler }data base led to several conclusions.

\begin{description}
\item[(1)] The configurations of all targets are almost the same: 4 of them are contact binaries (KID~4921906, KID~1572353, KID
8108785, KID~6309193) and 2 are slightly overcontact systems (KID~12055255 and KID~9532219).
\item[(2)] Two targets, KID~4921906 and KID~6309193, are noneclipsing systems. The registration of their low-amplitude ellipsoidal and rotational 
(due spots) variability is possible due to the high \textit{Kepler} accuracy. There are quite many objects of this type in the \textit{Kepler} EB 
catalog and hence, it should be called catalog of \textit{photometrically-double} stars.
\item[(3)] To model the data of KID~9532219 we had to assume considerable contribution of a third light $l_3$. It originates probably from
the close neighbor. We do not exclude this fainter star to be the true variable because the size of the \textit{Kepler} pixels projected onto the sky 
(4 arcsec) coupled with the high star density near the Galactic plane lead to a non-negligible likelihood of associating an EB event with the wrong 
star (PRSA11).
\item[(4)] The target KID~12055255 is very rare case of ultrashort-period binary consisting of two M dwarfs. The system seems to be overcontact. 
We are not aware of any overcontact binary consisting of M dwarfs.
\item[(5)] Feldmeier et al. (2011) classified the target KID~9532219 as a $\delta$ Sct variable (from their ground-based photometric observations 
of the \textit{Kepler} FoV for the pre-launch survey). The \textit{Kepler} data and our modeling certainly reveal that this target (or its neighbor) 
is eclipsing binary.
\item[(6)] Recently KID~6309193 was noted as "false" in the \emph{Kepler} EB catalog. Our modeling showed that it is not eclipsing star but
its light curve can be reproduced by ellipsoidal variations of binary. However we do not exclude the alternative the observed variability to be 
inherent to some of its two considerably fainter neighbors (at distances around 4.5 arcsec and 7 arcsec).
\item[(7)] The ripples in the residuals of KID~8108785 (Fig. 3) seem to be quasi-sinusoidal with period of 55 min and amplitude of 0.005
mag. Although this variability is visible during all phases, it could be explained by inherent changes of some of the components
of this system due to its small orbital inclination.
\end{description}

Although all considered light curves reveal O'Connell effect and are reproduced by cool spots we have not studied long-term light curve changes as a 
result of the evolution of these surface inhomogeneities. To do this one should have detailed information about the problems of the satellite 
steering which introduce different trends into the data.

\section{Conclusions}

Our investigation added six new members to the small family of the studied ultrashort-period binaries with MS components.

All six targets turned out to be contact or slightly overcontact configurations which components are almost in thermal contact. Two targets, 
KID~4921906 and KID~6309193, are not eclipsing but reveal ellipsoidal and spot variability. KID~12055255 is very rare case of ultrashort-period 
binary consisting of two M dwarfs. We established that KID~9532219 is an eclipsing binary but not a $\delta$ Sct variable as it was previously 
supposed. One of the components of KID~8108785 exhibits inherent, quasi-sinusoidal, small-amplitude variability.

The analysis of the results showed that reliable global parameters of the \textit{Kepler} binaries might be obtained only by standard method of 
light curve solutions. The automated \textit{EBAI} method models the \textit{Kepler} data quite roughly. Obviously, the numerous and exclusive 
precise \textit{Kepler} data deserve precise light curve solutions by the standard way.

\acknowledgements We used the live version of the \textit{Kepler} EB catalog from http://keplerEBs.villanova.edu. The research was supported partly by 
funds of project RD 08-244 of Scientific Foundation od Shumen University. This publication makes use of data products from the Two Micron All Sky 
Survey, which is a joint project of the University of Massachusetts and the Infrared Processing and Analysis Center/California Institute of 
Technology, funded by the National Aeronautics and Space Administration and the National Science Foundation. This research also has made use of the 
USNOFS Image and Catalogue Archive operated by the United States Naval Observatory, Flagstaff Station (http://www.nofs.navy.mil/data/fchpix/).

The authors are grateful to anonymous referee for the valuable notes and propositions.

\end{document}